\documentstyle[12pt]{article}

\def\be{\begin{equation}}
\def\ee{\end{equation}}

\def\e{\epsilon}

\def\vol{\hat{\rm vol}}

\def\vol{\widehat{{\rm vol}}}

\def\Pexp{{\rm Pexp}}

\def\Cyl{{\rm Cyl}}

\def\be{\begin{equation}} 
\def\ee{\end{equation}} 
\def\ba{\begin{eqnarray}} 
\def\ea{\end{eqnarray}}

\def\A{{\cal A}}

\def\S{\Sigma}\def\g{\gamma} 
 
\def\Ab{{\overline \A}}

\def\g{\gamma} 
 
\def\Comp{{\mathchoice 
{\setbox0=\hbox{$\displaystyle\rm C$}\hbox{\hbox to0pt 
{\kern0.4\wd0\vrule height0.9\ht0\hss}\box0}} 
{\setbox0=\hbox{$\textstyle\rm C$}\hbox{\hbox to0pt 
{\kern0.4\wd0\vrule height0.9\ht0\hss}\box0}} 
{\setbox0=\hbox{$\scriptstyle\rm C$}\hbox{\hbox to0pt 
{\kern0.4\wd0\vrule height0.9\ht0\hss}\box0}} 
{\setbox0=\hbox{$\scriptscriptstyle\rm C$}\hbox{\hbox to0pt 
{\kern0.4\wd0\vrule height0.9\ht0\hss}\box0}}}} 
\def\Co{{\mathchoice 
{\setbox0=\hbox{$\displaystyle\rm C$}\hbox{\hbox to0pt 
{\kern0.4\wd0\vrule height0.9\ht0\hss}\box0}} 
{\setbox0=\hbox{$\textstyle\rm C$}\hbox{\hbox to0pt 
{\kern0.4\wd0\vrule height0.9\ht0\hss}\box0}} 
{\setbox0=\hbox{$\scriptstyle\rm C$}\hbox{\hbox to0pt 
{\kern0.4\wd0\vrule height0.9\ht0\hss}\box0}} 
{\setbox0=\hbox{$\scriptscriptstyle\rm C$}\hbox{\hbox to0pt 
{\kern0.4\wd0\vrule height0.9\ht0\hss}\box0}}}} 
\def\Rl{{\mathchoice 
{\setbox0=\hbox{$\displaystyle\rm R$}\hbox{\hbox to0pt 
{\kern0.4\wd0\vrule height0.9\ht0\hss}\box0}} 
{\setbox0=\hbox{$\textstyle\rm R$}\hbox{\hbox to0pt 
{\kern0.4\wd0\vrule height0.9\ht0\hss}\box0}} 
{\setbox0=\hbox{$\scriptstyle\rm R$}\hbox{\hbox to0pt 
{\kern0.4\wd0\vrule height0.9\ht0\hss}\box0}} 
{\setbox0=\hbox{$\scriptscriptstyle\rm R$}\hbox{\hbox to0pt 
{\kern0.4\wd0\vrule height0.9\ht0\hss}\box0}}}} 
\def\L{{\rm L}}

\def\L2{L^2(\Ab, d\mu_0)}

\def\-{^{\rm down}}
\def\+{^{\rm up}} 

\begin{document} 
\title{Volume and Quantizations} 
\author{Jerzy Lewandowski
\thanks{Institute of Theoretical Physics,
 Warsaw University, ul Hoza 69, 00-681 Warsaw, Poland
 and
 Max-Planck-Institut f\"ur Gravitationsphysik,
 Schlaatzweg 1, 14473 Potsdam, Germany}}

\maketitle

\begin{abstract}
The aim of this letter is to indicate the differences between the
Rovelli-Smolin quantum volume operator and other quantum volume
operators existing in the literature. The formulas for the operators
are written in a unifying notation of the graph projective framework.
It is clarified whose results apply to which operators and 
why.  
\end{abstract} 
\eject
\newpage
It should be emphasized at the very beginning that the letter  
has been motivated by  very nice calculations made recently  by Loll 
\cite{Loll1} in connection with the lattice quantization of the volume and 
by the work of Pietri and Rovelli \cite{PR}  who studied  the Rovelli-Smolin
volume operator \cite{Rovelli95}. In the full continuum theory, there is 
still one 
more candidate for the quantum volume operator proposed in \cite{ALg,L,A}
which  will be referred to as the 
`projective limit framework'
volume operator. The misunderstanding which we   indicate
here is that in \cite{Loll1,Loll2}  the lattice operator tends to be considered
 as the restriction of the Rovelli-Smolin as well as  the 
projective limit framework volume operators. In fact,
 the first lattice volume, that of \cite{Loll1}, corresponds to neither of 
them. 
The corrected lattice volume operator of \cite{Loll2}, on the other hand, 
has been
modified to agree with the graph projective framework volume operator.
However this is still different then the Rovelli-Smolin operator.    
We don't discuss here the origins of the differences between the two
continuum theory operators or compare their statuses. This is the subject 
of the coming paper \cite{ALv}. Therein, the Rovelli-Smolin operator
has been written in terms of the graph projective framework which makes
the comparison possible. Below we  present the derived formula
and show the difference between the Rovelli-Smolin and
the graph projective framework  operators. 
Finally, we explain why some of the arguments of Loll are still true
for the both remaining operators and what is the origin of
the relations between some eigen values derived by Loll and
those established  by  Pietri and Rovelli. 
 
\bigskip 

All the operators in question come from the same classic formula
for the volume of a 3-surface $\Sigma$ of initial data expressed by 
the Ashtekar variables,  
\be
{\rm vol}\ =\ \int_\S d^3x \sqrt{{1\over 3!}|\e^{ijk}\e_{abc}{E_i^a(x)
E_j^b(x)E_k^c(x)}|}, 
\ee
where $E^a_i$ is the $su(2)$ valued vector-density momentum canonically
conjugate to the Ashtekar connection $A^i_a$, $a,b,c$ being the space
and $i,j,k$ the internal indices. The space of the connections 
is denoted by $\A$. 
 
The Hilbert space in which we will define the operators is constructed  from 
the gauge invariant cylindrical functions on $\A$  
\cite{1,2,3,4,ALg}. 
Recall that a function $\Psi$ defined on  $\A$ is called {\it cylindrical} if 
there exists a finite family  $\g =\{e_1,...,e_n\}$ of finite curves 
and a complex valued  function $\psi\in C^0(SU(2)^n)$ such that
\be \label{cyl}
\Psi(A)\ =\ \psi(U_{e_1}(A),...,U_{e_n}(A)),\ \ \ U_p\ =\ \Pexp{\int_p -A};
\ee
that is, where  given a curve $p$ in $\S$,  $U_p$ is the parallel 
transport matrix with respect to $A$. If we admit only piecewise analytic
curves then for every cylindrical function we can choose $\g=\{e_1,...,e_n\}$ 
to be an 
embedded graph, meaning that every curve (edge) $e_i$ is an analytic 
embedding   of an interval and two distinct elements of $\g$ can share, 
if any,  only   one or the both end points (called the vertices 
of $\g$). The Hilbert space is the space $L^2(\Ab))$ of cylindrical functions
integrable with the square with respect to the following integral
\be
\int_\A \Psi\ =\ \int_{SU(2)^n}\psi(g_1,...,g_n)dg_1...dg_n
\ee  
$dg$ being the probability Haar measure on $SU(2)$. 

The volume operator involves third derivatives, so for its initial domain 
(later on extended by the essential self-adjointness) we  will take the 
space $\Cyl^{(3)}$ of the  cylindrical functions given by the $C^3$ 
functions $\psi$.  
 
Before writing the explicit formulas we need a `basis'  of 
first order differential operators acting on the cylindrical functions. 
To a pair, an analytic curve $p$ and its marked endpoint  $v$,
and to a an element $\tau_i$ of a fixed basis in $su(2)$ we            
assign an operator $X_{vpi}$ acting on cylindrical functions
as follows. Given a cylindrical function, represent it by (\ref{cyl})
using a graph such that one of its edges, say $e_I$, is a segment of $p$
containing the point $v$. Then, respectively
\be
X_{vpi}\Psi(A)\ :=\ \cases{\big(U_{e_I}\tau_i\big)^A_B{\partial\over
\partial {U_{e_I}}^A_B}\psi(U_{e_1}(A),...,U_{e_n}(A)),& \cr
\big(-\tau_iU_{e_I}\big)^A_B{\partial\over\partial {U_{e_I}}^A_B}
\psi(U_{e_1}(A),
...,U_{e_n}(A)),&\cr}                             
\ee 
when $e_I$ is outgoing and  when $e_I$ is incoming, 
at the vertex $v$ (the result is graph independent). 

The `graph projective framework' volume operator of \cite{ALg,L,A}
acts on a cylindrical function (\ref{cyl}) in the following way
(below, we skip an overal constant factor which is the same for all the 
operators)

\be\label{v}
\vol\, \Psi\ =\ \sum_v\sqrt{|\hat{q}_v|}\Psi
\ee
where the sum ranges over all the vertices in the graph $\g$ used 
to represent $\Psi$ by (\ref{cyl}) and the operator  $\hat{q}_v$
assigned to a point $v$ of $\S$ is defined by
\be\label{ALq}
\hat{q}_v\Psi\ =\ {1\over 8\cdot 3!} \sum_{(e_I,e_J,e_K)} i
\e(e_I,e_J,e_K)\e_{ijk}X_{ve_Ii}X_{ve_Jj}X_{ve_Kk},
\ee
the sum ranging over all the ordered triples of edges of $\g$  at $v$
and where $\e(e_I,e_J,e_K)$ depends only on the orientation
of the vectors $(\dot{e}_I,  \dot{e}_J, \dot{e}_K)$ at $v$,
\be
\e(e_I,e_J,e_K)\ =\ \cases{1,& when the orientation is positive,\cr
                           -1,& when negative,\cr
                            0,&when the vectors are linearly dependent.\cr} 
\ee

On the other hand, the Rovelli-Smolin volume regularization \cite{Rovelli95} 
written in terms of the graph projective framework turns out
to give the following result \cite{ALv}
\be\label{RS}
\vol^{\rm RS}\, \Psi\ =\ \sum_v\sqrt{\hat{q}^{\rm RS}_v}\Psi
\ee
where
\be\label{RSq}
\hat{q}^{\rm RS}_v\Psi\ =\ {1\over 8\cdot 3!} \sum_{(e_I,e_J,e_K)} |\e_{ijk}
X_{ve_Ii}X_{ve_Jj}X_{ve_Kk}|
\ee
using the same sum convention as above. (Surprisingly, the factor
$8$ in (\ref{ALq}) appears on the quantum level whereas in (\ref{RS})
it comes from the classical expression used to approach the determinant.)

To understand the difference between $\vol$ and $\vol^{\rm RS}$, 
notice that  the  volume operator $\vol$
is sensitive on {\it diffeomorphic characteristics} of intersections in
a given graph. For a planar graph for instance each $\hat{q}_v$ is 
identically zero.
On the other hand, the Rovelli-Smolin operator $\hat{q}^{\rm RS}_v$
extracted by our decomposition, regards each triple
of edges at an intersection equally,  irrespectively whether they are tangent
or not. Thus, whereas the `graph projective framework' volume $\vol$
is diffeomorphism of $\S$ invariant, the Rovelli-Smolin
volume  $\vol^{\rm RS}$   is preserved by all the homeomorphisms!
To be more precise: given a cylindrical function $\Psi$ of (\ref{cyl}), 
if a diffeomorphism (homeomorphism) $\varphi$ of $\S$ happens to
carry a graph $\g$ representing $\Psi$ into a
graph which {\it is again}   piecewise analytic, then the induced action of 
$\varphi$ on $\Psi$ commutes
with the action of $\vol$ (respectively, with $\vol^{\rm RS}$).

Let us turn now to the Loll graph operators. Now, a graph $\g$ is
a fixed cubic lattice.  The first operator published  in \cite{Loll1} is 
given by the Eqs (\ref{v}, \ref{ALq}) however with the sum in (\ref{ALq}) 
ranging only over the outgoing edges (with respect to an orientation
of the lattice) and without the factor ${1\over 8}$. Finally,  the modified
lattice volume operator of \cite{Loll2} coincides with the `graph projective 
framework' volume operator (\ref{v},\ref{ALq}) (modulo some mistakes: taken 
literally as it stands in \cite{Loll2}, that definition would give 
products of operators $X_{vei}X_{v'e'i}...$ at different points 
$v\not=v'$ which wouldn't be gauge invariant; but a simple correction 
corresponding to the previous verbal description removes this problem.)
Let us denote these operators by $\vol^{\rm L1}$ and $\vol^{\rm L2}$. 

Studying the operator $\vol^{\rm L1}$, Loll proved that it annihilates all
the cylindrical functions on a lattice which at each vertex involve not
more then three intersecting edges. Remarkably, this result continues
to be true for any of the remaining volume operators \cite{L,Loll2,PR}.
Moreover, recently the eigen values of $\vol^{\rm L2}$ 
 for four valent graphs embedded in
a lattice   (applicable 
to the operator $\vol$ automatically) obtained in \cite{Loll2} 
 are being compared and confirmed by the team studying the original    
Rovelli-Smolin volume operator $\vol^{\rm RS}$. Let us take advantage
of our unified notation to see  how it could  happen. 

In  3-valent case, at a vertex $v$ there are at most three edges.
 Thus the formulas (\ref{v},\ref{ALq}) and 
(\ref{RS},\ref{RSq}) just coincide. Moreover, every term
 $X_{ve_Ii}X_{ve_Jj}X_{ve_Kk}\e_{ijk}$ kills a gauge invariant cylindrical 
function individually \cite{Loll1}. We show this by an alternative calculation 
\cite{L}. Fix a vertex $v$ of a graph $\g$  and denote the edges at $v$
by  $e_1, e_2, e_3$ say. Restricted to the gauge invariant cylindrical
functions, the differential operators satisfy the following constraint,
(we drop $v$)
\be
 X_{e_1i}\ +\  X_{e_2i}\ +\  X_{e_3i}.\ =\ 0
 \ee
>From that we evaluate
\be
\epsilon_{ijk} X_{e_1i} X_{e_2j}X_{e_3k}\ =\ 
-\epsilon_{ijk} X_{e_1i} X_{e_2j} (X_{e_1k} + X_{e_2k}) \ = \ 0.
\ee

Consider now a four valent case. Let $e_1,...,e_4$ be the edges of $\g$
incident at a fixed vertex $v$. Then,  the Gauss constraint reads
\be
  X_{e_1i}\ +\  X_{e_2i}\ +\  X_{e_3i}\ +\  X_{e_4i}\ =\ 0.
\ee
>From that we derive 
\be
\epsilon_{ijk} X_{e_1i} X_{e_2j}X_{e_3k}\ =\ 
-\epsilon_{ijk} X_{e_1i} X_{e_2j}X_{e_4k}
\ee
etc.. That is, at a vertex, all the triple products 
 $X_{ve_Ii}X_{ve_Jj}X_{ve_Kk}\e_{ijk}$  are equal modulo a sign.
Hence, for a four valent case, at every vertex
\be
 \hat{q}_v\ =\ \kappa(v)  \hat{q}^{\rm RS}_v,
\ee
the constant factor being given by the diffeomorphic
versus topological characteristics of the intersection at $v$. 
This explains, why the eigen values  of \cite{Loll2} can be in a direct 
relation
with the results of of \cite{PR}   even in the 4-valent case as long as
 a single vertex is considered.

Since the Rovelli-Smolin operator $\vol^{\rm RS}$ is presented here for 
the first
time in the above form  let us
briefly explain how in the graph projective framework one can see its 
self adjointness
(as well as the self adjointness of the operator $\vol$).
For every fixed graph $\g$ the operators   $iX_{ve_Ii}X_{ve_Jj}
X_{ve_Kk}\e_{ijk}$  are essentially-self adjoint
differential operators  acting in the domain $C^3(SU(2)^n)$ in  $L^2(SU(2)^n)$
(which justifies  taking the absolute value).
>From this one proves, that in the terminology of \cite{ALg} the operator
$\vol^{\rm RS}$ defines   a consistent family
of essentially self-adjoint operators. Therefore, as proven therein, it is
essentially self adjoint in $\Cyl^{(3)}$.       

To show the discreteness of the spectrum of $\vol^{\rm RS}$ 
(proven in \cite{PR})  and of the operator $\vol$, it is enough
to prove the discreteness of the operators reduced to any graph and
the corresponding $L^2(SU(2)^n)$ \cite{ALv}. For a fixed graph $\g$
we can use just the spin-network functions \cite{Rovelli95a,6}
(we are using the notation 
of \cite{7}). The operators  $X_{ve_Ii}X_{ve_Jj}
X_{ve_Kk}\e_{ijk}$  acting on the spin-network functions
$T^{\g,\vec{\pi},\vec{c}}$ preserve the labeling $\vec{\pi}$  of the edges 
of $\g$ by irreducible representations and only affect in a pointwise way  
the intertwining operators assigned to the vertices by $\vec{c}$ 
\cite{Rovelli95,Loll1,L}.
Thus they restrict to finite dimensional real and antisymmetric  matrices
which completes the proof.

 \bigskip
{\centerline {\bf Acknowledgments}}
It is a pleasure to thank Abhay Ashtekar for many hours of joint 
calculations and discussions when I was visiting 
the Center for
Gravitational Physics and Geometry of the Penn State University.  
Discussions with John Baez, Bernd Bruegman, Renate Loll, Don Marolf, Jose 
Mourao, Carlo Rovelli, Lee Smolin and Thomas Thiemann are
gratefully acknowledged. 
I would like to thank the members of the Max-Planck-Institut, where this
work was done,  for their hospitality. This work was supported in part by 
 the KBN grant 2-P302 11207

\end{document}